\documentstyle[11pt]{article}
\def\be{\begin{equation}}
\def\ee{\end{equation}}
\def\bea{\begin{eqnarray}}
\def\eea{\end{eqnarray}}
\newcommand{\atil}{\tilde a }
\newcommand{\ctil}{\tilde c }
\newcommand{\hip}{\, {}_1F_1 }

\newcommand{\undera}{{\underline a } }
\newcommand{\underc}{{\underline c } }

\begin{document}
\baselineskip=15pt
\def\harr#1#2{\smash{\mathop{\hbox to
.5in{\rightarrowfill}}\limits^{\scriptstyle#1}_{\scriptstyle#2}}}

\def\harrl#1#2{\smash{\mathop{\hbox to
.5in{\leftarrowfill}}\limits^{\scriptstyle#1}_{\scriptstyle#2}}}

\def\varr#1#2{\llap{$\scriptstyle #1$}\left\downarrow \vcenter to .5in
{}\right.\rlap{$\scriptstyle #2$}}  

\def\varru#1#2{\llap{$\scriptstyle #1$}\left\uparrow \vcenter to .5in
{}\right.\rlap{$\scriptstyle #2$}}

\vspace*{-2.5cm}
\begin{center}
{\tiny Proceedings of the {\em Fourth Mexican School on Gravitation
and Mathematical Physics} ``Membranes 2000''. Huatulco O\'ax. 
Mexico. N. Bret\'on, J.S. D\'{\i}as and H. Quevedo (Eds)}
\end{center}

\begin{center}
{\LARGE{\bf Physical sectors of the confluent hypergeometric}}
\smallskip

{\LARGE{\bf functions space}}
\end{center}
\vskip0.5cm

\begin{center}
Oscar Rosas-Ortiz\\[0.1cm]
{\em
Departamento de F\'\i sica, CINVESTAV-IPN\\
A.P. 14-740, 07000 M\'exico D.F., Mexico
}
\end{center}  

\begin{center}
J. Negro and L.M.~Nieto\\[0.1cm]
{\em
Departamento de F\'\i sica Te\'orica, Universidad de Valladolid\\
47011 Valladolid, Spain
}
\end{center}  
\vskip1cm

\begin{abstract}
A relaxed factorization is used to obtain many of the properties
obeyed by the confluent hypergeometric functions. Their implications
on the analytical solutions of some interesting physical problems are
also studied. It is quite remarkable that, although these properties
appear frequently in solving the Schr\"odinger equation, it has been
not clear the role they play in describing the physical systems. The
main objective of this communication is precisely to throw some light
on the subject. 
\end{abstract}

\section{Introduction}

There is no doubt on the main role played by special functions in
theoretical and mathematical physics. In general, they are used to
simplify the original problem by transforming its mathematical
description from a rather cumbersome form into a simpler and well
known one. The physical scenery is thereby clarified: the solutions
of the simplified problem can be easily analized and their most
relevant qualitative features could be depicted in terms of the
involved parameters. As a matter of fact, a considerable amount of
basic research has been developed in the study of the differential
equations obeyed by special functions, mainly in their connections
with the problems appearing in all the branches of the physical
theories. Of particular interest are the hypergeometric and confluent
hypergeometric functions (h.f. and c.h.f.  respectively), in terms of
which almost all the solutions of exactly solvable problems in
quantum mechanics can be written (e.g. those related with the linear
and p-dimensional harmonic oscillator, hydrogen-like,
P\"oschl-Teller, Wood-Saxon, Hulth\'en, Morse, Eckart and Scarf
potentials among others).  The standard mechanism in these cases
takes into account appropriate transformations of the involved
variables and functions from the Schr\"odinger into an hypergeometric
or confluent hypergeometric equation (h.e. and c.h.e. for the last
two respectively) \cite{Flu71}. Remark on the fact that the
mathematical properties of the h.f. and c.h.f. lead then to the
characterization of some important physical features of the
solutions. That is the case, for example, of the quantization of the
energy eigenvalues by imposing the Schr\"odinger solutions of
classically confined systems as square integrable functions. On the
other hand, an interesting perspective arises by considering the
procedure in the reverse order: one departs from the h.e. or the
c.h.e. and the conditioning transformations leading to the
Schr\"odinger equation are to be determined (see important works by
Natanzon \cite{Nat71} and by Nikiforov and Uvarov \cite{Nik88}). Such
a procedure allows the identification of a wide set of functions that
can be understood as physically meaningful potentials.

Considering all this theoretical richness, it seems apparent the
presence of relations connecting the stationary Schr\"odinger
solutions of diverse quantum systems. Although they have been
frequently mentioned in the literature since long time ago, there is
still lacking an updated exposition of the topic. Quite recently, the
present authors have reported results which could shed a new light on
this subject \cite{Neg00}. In that work a refined factorization was
applied to analize the c.h.e. and its solutions, as well as the
implications of their mathematical properties on the wave-functions
of some interesting physical problems. The refined factorization goes
deeply into the possibilities of the conventional factorization
method providing with additional significative information
\cite{Neg00b}. 

In this contribution we shall discuss some of the main results
published in \cite{Neg00} whereas it is also reported more on their
physical consequences. The Section 2 is devoted to the refined
factorization of the c.h.e. The action of the involved factorization
operators on the confluent hypergeometric parameters underlies the
analytical and algebraic properties of the c.h.f space. Section 3
deals with the mapping from the c.h.e. into diverse Schr\"odinger
equations. The identification of the involved conventional physical
potentials gives the chance to translate the results obtained in
Section 2 from a mathematical into a physical language.

\section{Confluent hypergeometric intertwiners}

We shall work on second order differential operators of the form
\be
L_{(a,c)} \equiv x \frac{d^2}{dx^2} + (c-x) \frac{d}{dx} -a
\label{uno}
\ee
where the pair $(a,c)$ represents a point on ${\bf R}^2$. Once the
values of $a$ and $c$ have been given, the kernel elements of
$L_{(a,c)}$ can be characterized as c.h.f., {\it i.e.\/}, $f(a,c;x)$
is a c.h.f. iff $L_{(a,c)} f(a,c;x)=0$, or in other words, $f(a,c;x) 
\in {\cal K}_{(a,c)}$. Thereby, one can look for a way to connect the
kernel ${\cal K}_{(a,c)}$ with another one ${\cal K}_{(\atil,\ctil)}$
and the conditions relating their respective parameters $(a,c)$ and
$(\atil, \ctil)$. With this aim let us consider an arbitrary
differential operator $X$, defined by its action on a kernel element
$f(a,c;x)$ of $L_{(a,c)}$, as follows
\be
X f(a,c;x):= X_{(a,c)} f(a,c;x)= \xi(\bar a, \bar c;x)
\label{x}
\ee
where, according with our earlier convention, $\xi(\bar a, \bar c;x)$
is a c.h.f. iff $\xi(\bar a, \bar c;x) \in {\cal K}_{(\bar a, \bar
c)}$ and the pair $(\bar a, \bar c)$ depends on $a$ and $c$. We shall
say that the operator $X$ is in a free index notation whereas it is
written in the confluent hypergeometric notation $X_{(a,c)}$ in
dependence on the kernel ${\cal K}_{(a,c)}$ on which it is acting.
The benefits of this notation become clear in the composition of any
pair $X, Y$ of these operators. For instance, if $Y$ is such that
${Y} f(a,c;x)= g(\atil,\ctil;x)$, then the action of $XY$ on the
kernel ${\cal K}_{(a,c)}$ reads
\[
( X Y) f(a, c;x) := { X}_{(\atil,\ctil)} \left( {Y}_{(a,c)} f(a,
c;x)\right) = { X}_{(\atil,\ctil)} g(\atil, \ctil;x).
\]
On the other hand, the action of $YX$ on the same kernel gives
\[
(YX) f(a, c;x) := Y_{(\bar a, \bar c)} \left( X_{(a,c)} f(a,c;x) \right)=
Y_{(\bar a, \bar c)} \xi(\bar a, \bar c;x).
\]
Taking full adavantage of our considerations we look now for a couple of
differential operators $A$ and $B$ such that
\be
A: {\cal K}_{(a,c)} \mapsto {\cal K}_{(\atil,\ctil)}, \qquad 
B: {\cal K}_{(\atil,\ctil)} \mapsto {\cal K}_{(a,c)}.
\label{intertwin}
\ee
The initial parameters $(a,c)$ can also play the role of being the
final ones. In that case we use the parameters $(\undera, \underc)$,
\be
A: {\cal K}_{(\undera, \underc)} \mapsto {\cal K}_{(a,c)}, \qquad
B: {\cal K}_{(a,c)} \mapsto {\cal K}_{(\undera, \underc)}.
\ee
It is a matter of substitution to verify that operators $A$ and $B$
intertwin the elements of ${\cal K}_{(a,c)}$ with those of ${\cal
K}_{(\atil,\ctil)}$ and vice versa. Table~1 displays the set of basic
intertwiners of diverse orders, all of them written in the confluent
hypergeometric notation.

\vskip-0.3cm

\begin{table}[!h]
\caption{\footnotesize Diverse order intertwiners for the c.h.o. 
$R_x$ is the reflection operator $R_x \varphi (x) = \varphi (-x)$.}

{\tiny
\begin{tabular*}{\textwidth}{@{\extracolsep{\fill}}lllll}
\hline\\
&&&&\\[-4ex]
Order & Intertwiner & Expression & $\quad\atil $ & $\quad\ctil$ 
\\
&&&&\\[-2ex]
\hline\\
&&&&\\[-3ex]
zero & $Q_{(a,c)}$ & $x^{c-1}$ & $a-c+1$ & $2-c$ 
\\ [2ex]
\hline\\
First & $A^1_{(a,c)}$ & $\frac{d}{dx}-1$ & $a$ & $c+1$ 
\\[2ex]
First & $B^1_{(a,c)}$ & $ x \frac{d}{dx}+c-1$ & $a$ & $c-1$ 
\\[2ex]
First & $A^2_{(a,c)}$ & $x \left(\frac{d}{dx}-1 \right) +c -1$ &
$a-1$ & $c-1$
\\[2ex]
First & $B^2_{(a,c)}$ & $\frac{d}{dx}$ & $a+1$ & $c+1$
\\[2ex]
First & $A^3_{(a,c)} $& $x^c\left( \frac{d}{dx}-1 \right)$ & $a-c$ &
$1-c$
\\[2ex]
First & $B^3_{(a,c)}$ & $ x^c \frac{d}{dx}$ & $a-c+1$ & $1-c$
\\[2ex]
First & $A^4_{(a,c)}$ & $x^{c -2} \left[ x \left( \frac{d}{dx}-1
\right) +c -1 \right]$ & $a-c+1$ & $3-c$
\\[2ex]
First & $B^4_{(a,c)}$ & $ x^{c-2} \left(x\frac{d}{dx}+c-1 \right)$ &
$a-c+2$ & $3-c$
\\ [2ex]
\hline\\
non-finite & $V_{(a,c)}$ & $e^x R_x$ & $c-a$ & $c$ 
\\ [2ex]
non-finite & $W_{(a,c)}$ & $x^{c-1}e^x R_x$ & $1-a$ & $2-c$ \\
&&\\[-2ex]
\hline
\end{tabular*}
}
\end{table}

It is remarkable the presence of the zero order differential operator
$Q$, which gives rise to a nontrivial intertwining operation:
$L_{(a-c+1,2-c)} Q_{(a,c)} = Q_{(a,c)} L_{(a,c)}$. On the other hand,
the non-finite order intertwiners $V$ and $W$ underlies some of the
very basic relations obeyed by the c.h.f. such as the well known {\em
Kummer's first formula} \cite{Neg00}. The first order intertwiners
$\{ A^i, B^i\}^4_{i=1}$ deserve as much attention because they
factorize the c.h.o.  $L_{(a,c)}$ and $L_{(\atil, \ctil)}$: 
\bea
\label{factor1}
L_{(a,c)}= B^i_{(\atil,\ctil)} A^i_{(a,c)} -q^i_{(a,c)}; \qquad
L_{(a,c)}= A^i_{(\undera, \underc)} B^i_{(a,c)} -q^i_{(\undera,\underc)}, 
\qquad  i=1,2,3,4.\\
L_{(\atil,\ctil)}= A^i_{(a,c)} B^i_{(\atil,\ctil)} -q^i_{(a,c)}; \qquad
L_{(\undera,\underc)}= B^i_{(a,c)} A^i_{(\undera,\underc)} 
-q^i_{(\undera, \underc)}, \qquad  i=1,2,3,4.
\label{factor2}
\eea
where, by convention, we have implemented the {\em induced action} of
the operators $A$ and $B$ on ${\bf R}^2$ as follows:  $A^i(a,c):=
A^i_{(a,c)} (a,c)= (\atil,\ctil)$, $B^i (a,c):= B^i_{(a,c)}
(a,c):=(\undera, \underc)$. The factorization constants $q^i_{(a,c)}$
are given by
\[
q^1_{(a,c)}= a-c, \quad q^2_{(a,c)}=a-1, \quad 
q^3_{(a,c)}=a-c, \quad q^4_{(a,c)}=a-1.
\]
Now, in order to determine the explicit action of the free index
operators on an arbitrary kernel ${\cal K}_{(a,c)}$, it is enough to
consider a basis for ${\cal K}_{(a,c)}$ as follows
\be
{}_1F_1 (a,c;x), \qquad u(a,c;x) \equiv x^{1-c} {}_1F_1(a-c+1,2-c;x),
\label{basis}
\ee
where it is assumed that $c \notin {\bf Z}$ (the general case $c \in {\bf
Z}$ is easily recovered afther an straightforward limiting procedure
\cite{Neg00}). The action of the intertwiners $Q,V,W$ and $\{ A^i,
B^i\}_{i=1}^4$ on this basis is displayed on Table~2. 

\vskip-0.3cm

\begin{table}[!hbp]
\caption{\footnotesize Action of the basic free index intertwiners on
the basis of ${\cal K}_{(a,c)}$.}

{\tiny
\begin{tabular*}{\textwidth}{@{\extracolsep{\fill}}cll}
\hline
\\ &&\\[-4ex]
Operator & $\quad \quad \hip (a,c;x)$ & $\quad \quad u(a,c;x)$ 
\\ &&\\[-2ex]
\hline
\\ &&\\[-4ex]
$Q$ & $u (a+1-c, 2-c; x)$ \qquad & $\hip
(a+1-c, 2-c; 
x)$ 
\\ [2ex]
\hline\\
$A^1$ & $\left( \frac{a-c}{c} \right)
\hip(a,c+1;x)$ & $(1-c) \, u(a,c+1;x)$ 
\\ [2ex]
$B^1$ & $(c-1) \hip(a,c-1;x)$ & $\left(
\frac{a-c+1}{2-c} \right) u(a,c-1;x)$ 
\\ [2ex]
$A^2$ &
$(c-1) \hip (a-1,c-1;x)$ & $\left( \frac{a-1}{2-c} \right)\,
u(a-1,c-1;x)$ 
\\ [2ex]
$B^2$ & $\left(
\frac{a}{c} \right) \hip (a+1, c+1, x)$ & $(1-c) \, u(a+1,c+1;x)$
\\ &&\\[-2ex]
$A^3$ & $
\left(
\frac{a-c}{c}
\right) \, u(a-c,1-c;x)$ & $(1-c) \hip (a-c,1-c;x)$
\\ [2ex]
$B^3$ & $ \left( \frac{a}{c} \right) \,
u(a-c+1,1-c;x)$ & $(1-c) \hip (a-c+1,1-c;x)$
\\ [2ex]
$A^4$ &  $(c-1) \, u(a-c+1, 3-c, x)$ & $\left(
\frac{a-1}{2-c} \right)  \hip (a-c+1,3-c;x)$
\\ [2ex]
$B^4$ & $(c-1) \, u(a-c+2,3-c;x)$ & $\left(
\frac{a-c+1}{2-c} \right) \hip (a-c+2,3-c;x)$ 
\\[2ex]
\hline\\
$V$ & $\hip (c-a, c; x)$ & $(-1)^{1-c} u (c-a,
c; x)$
\\ [2ex]
$W$ \qquad& $u (1-a, 2-c; x)$ &
$(-1)^{1-c}
\hip (1-a, 2-c;x)$
\\ &&\\[-2ex]
\hline
\end{tabular*}
}
\end{table}

As regards the composition of intertwiners, we first notice that
$Q^2=1$. Hence, the kernel ${\cal K}_{(a,c)}$ is invariant under the
twice iterated action of $Q$. The other operators are interrelated by
means of $Q$: 
\be
W=QV, \,
A^2 = Q A^1Q,\,
A^3=Q A^1,\,
A^4=Q A^2,\,
B^2 = Q B^1Q, \,
B^3=Q B^2,\,
B^4=Q B^1.
\label{productos}
\ee
These expressions can be now used to compute the commutation rules
obeyed by the intertwiners. On the other hand, according with
Table~2, the functions ${}_1F_1(a,c;x)$ and $u(a,c;x)$ are such that
they interchange roles under the action of $Q$. The same holds for
other operators such that $W$ or $A^4$ but adding a multiplicative
non trivial constant\footnote{This unpleasant situation can be
solved by introducing maximal shape-invariant functions preserving
their form under the action of all the intertwiners on Table~1
\cite{Neg00}.}. Thereby, $Q$ becomes an important intertwiner in our
approach, it works simply by transforming one element of the basis
into the other one and vice versa.  Such a behaviour is translated
from the c.h.f. space into the space of parameters as a reflection
operation. That statement is clear by considering a more
convenient parametrization of ${\bf R}^2$, defined by the
transformation
\be
a'=2a-c, \qquad c'=c-1, \qquad \forall \, a,c \in {\bf R}^2.
\label{transf}
\ee
With this new parametrization the operators $Q$ and $V$ are rewritten
in such a way that their induced action on ${\bf R}^2$ becomes
linear, homogeneous and diagonal: 
\be
Q(a',c')=(a',-c'), \qquad V(a',c')=(-a',c').
\label{refleja}
\ee
The induced action of the first order intertwiners in terms of the
new labels is also simpler
\be
\left\{
\begin{array}{l}
{ A}^1  (a',c') = (a'-1,c'+1) ,  \\ [1ex]
{ B}^1  (a',c') = (a'+1,c'-1) ,
\end{array} \right.
\quad
\left\{
\begin{array}{l}
{ A}^2  (a',c') = (a'-1,c'-1) , \\ [1ex]  
{ B}^2  (a',c') = (a'+1,c'+1) . 
\end{array} \right.
\label{a1b1}
\ee
\be  
\left\{
\begin{array}{l}
{ A}^3  (a',c') = (a'-1,-c'-1) ,  \\ [1ex]
{ B}^3  (a',c') = (a'+1,-c'-1) ,
\end{array} \right.
\ \
\left\{
\begin{array}{l}
{ A}^4  (a',c') = (a'-1,-c'+1),  \\ [1ex]
{ B}^4  (a',c')  = (a'+1,-c'+1) .
\end{array} \right.
\label{a3b3}
\ee
From (\ref{refleja}) and Table~1 we see that $V$ and $W$ are also
reflection operators in the parameter domain. A different situation
arises for the other intertwiners. For instance, according to
(\ref{a1b1}), the action of $A^1$ and $B^1$ on the point $(a',c') \in
{\bf R}^2$ produces the displacements of $a'$ and $c'$. Therefore, by
iterating the action of $A^1$ on ${\bf R}^2$ we obtain $(A^1)^n \,
(a',c') = (a'-n,c'+n)$, while for $B^1$ it reads $(B^1)^m \, (a',c') 
= (a'+m,c'-m)$. All these points are indeed in a straight line and
form a linear discrete lattice on the space of parameters: $\{
(a'+s,c'-s), s \in {\bf Z} \}$. For each one of these points we can
associate a kernel ${\cal K}_{(a'+s,c'-s)}$ and, in this way, the
action of $A^1$ or $B^1$ on a c.h.f. $f(a',c';x)$ can be understood
as the mapping from ${\cal K}_{(a',c')}$ into ${\cal K}_{(a'-1,
c'+1)}$ or ${\cal K}_{(a'+1, c'-1)}$ respectively. The iteration
procedure is now clear and the complete set of related c.h.f. $\{
f(a'+s, c'-s; x)\}_{s=-\infty}^{\infty}$ is nothing but an invariant
subspace under the action of $A^1$ and $B^1$. A similar situation
occurs for the other first order intertwiners. In general, when the
kernel of an operator $X^k \in \{A^k,B^k\}_{k=1}^{4}$ is in a kernel
${\cal K}_{(a',c')}$, the involved invariant subspace is bounded. We
shall call an {\em annihilation line} of the operator $X^k$, denoted
by $al[X^k]$, to the set of points $(a',c') \in {\bf R}^2$ such that
$f(a',c';x) \in {\cal K}_{(a',c')}$ iff $X^k_{(a',c')} f(a',c';x)=0$. 
From (\ref{factor1}) and (\ref{factor2}) it is easy to see that the
factorization lines for the operators $A^k$ and $B^k$ are given by
$q^k_{(a',c')}=0$ and $q^k_{(\undera',\underc')}=0$ respectively. 
After a simple calculation and using (\ref{productos}) one gets the
following relationships
\[
al[A^3]=al[A^1], \quad 
al[A^4]=al[A^2], \quad 
al[B^3]=al[B^2], \quad 
al[B^4]=al[B^1], 
\]
which give the chance to construct a common set of functions ${\cal
F} \equiv \{ f(a'_n, c'_m;x) \}$ representing an invariant space
under the action of $\{A^k, B^k\}_{k=1}^2$. In this case the points
$\{ (a'_n,c'_m)\}$ constitute a two dimensional discrete lattice on
${\bf R}^2$. The most interesting situation arises when there exists
a {\em critical point} on the lattice because, by definition, this point 
is shared by two different annihilation lines. A situation giving
rise to an {\em invariant sector} of ${\cal F}$, bounded by that lines. 
In fact, there are four critical points on our $(a',c')$
plane\footnote{The $a'$ values run on the horizontal axis in a right
hand frame of reference.}: $(-1,0)$, $(0,1)$, $(0,-1)$, and $(1,0)$. 
We can then define the following sectors: 

\noindent
{\bf Left Invariant Sector} ({\bf L.I.S.}). The point
$(a',c')=(-1,0)$ is at the intersection of $al [B^1]= \{(a',c') \,
\vert \, c'=a'+1 \}$ and $ al [B^2]=\{(a',c') \, \vert \, c'=-a'-1
\}$. A left sector of the lattice is then bounded by these
annihilation lines and it can be filled by successively applying
higher powers of $A^1$ and $A^2$ on $(-1,0)$. There is a {\em
left-corner} c.h.f. $f(-1,0;x) = 1$ connected with this point while
the related functions for the other points on the sector are of the
form
\[
f(-1-n,n;x) = (-1)^n, \, \, c' \geq 0; \qquad f(-1-m,-m;x) = (-x)^m, \,
\, c\leq 0. 
\]
In this way, we can characterize a support space generated by the
basis functions $\{f(-1-m-n,m-n;x),\ m,n\in {\bf Z}^+\}$, called
{\em left invariant sector} and denoted ${\cal H}_L$, for which there
is a (highest weight) irreducible representation of the algebra $\{{
A}^1,{ B}^1=-({ A}^1)^+,{ A}^2, { B}^2=-({ A}^2)^+,Q\}$ (for details
see \cite{Neg00}).

\noindent
{\bf Right Invariant Sector} ({\bf R.I.S.}). The critical point
$(a',c')=(1,0)$ is at the intersection of $ al [A^1]=\{(a',c') \,
\vert \, c'= a'-1\}$ and $al [A^2]=\{(a',c') \, \vert \, c'=
-a'+1\}$. The right sector bounded by these lines is generated by
means of ${ B}^1,{ B}^2$. The involved c.h.f. are given by
\[
g{(1+n,-n;x)} = x^n e^x, \, c' \leq 0; \qquad g{(1+m,m;x)} = e^x, \,
c' \geq 0,
\]
from which, the {\em right corner} function $g(1,0;x)=e^x$ can be
obtained by making $n=0$ or $m=0$. Therefore, we can construct a
(lowest weight) irreducible representation of the algebra $\{
A^1,B^1=-(A^1)^+, A^2,$ $B^2=-(A^2)^+,Q\}$, whose support space
spanned by $\{ g(1{+}n{+}m,m{-}n;x), n,m \in {\bf Z}^+ \}$ will be
denoted ${\cal H}_R$.

\noindent
{\bf Upper and Lower Sectors \/} The critical point $(0,-1)$ is at
the intersection of $al [A^3]$ and $al [B^3]$ while $(0,1)$ is at the
crossing of $al[A^4]$ and $al [B^4]$. However, there are not {\em
corner} c.h.f. associated with that points and one is unable to
construct another invariant sector (outside of the above mentioned
ones)  by means of any of the first order intertwiners $A^k$ or
$B^k$. Hence, there are only the two doubly bounded sectors already
mentioned.

It is now clear that only the left and the right sectors are of our
interest. It is remarkable that the intertwiner $Q$ becomes a self adjoint
operator in the support spaces related with each of these sectors.
Moreover, by taking a function $(-1)^n$ in the upper part of the left
sector we get, after the action of $Q$, a function $(-x)^n$ in the lower
part of that sector and vice versa. The same can be said about the
functions living in the right sector. Hence, this operator intertwins the
elements of a given sector with elements of the same sector. A very
different situation holds for the reflection operator $V$ which, by acting
on a function $(-x)^m$, gives $e^x x^m$. In other words, it maps the
functions living in the upper part of the left sector into the functions
of the upper part of the right sector and vice versa.  The same situation
occurs for the involved lower parts. The left and right invariant sectors
are then intertwined by means of the operators $Q$ and $V$ whereas they
are generated by the iterated action of the first order intertwiners on
the left and right corner functions. 

At a first sight, it seems to be enough to consider the points on the
lattice living in the left and right sectors. However, there is still
a subset of points deserving attention. The straight line defined by
the constraint $c'=-1/2$ results invariant under the action of $\{
A^3, B^3\}$. This line cuts $al[B^3]$ on $(-1/2,-1/2)$ and $al[A^3]$
on $(1/2,-1/2)$. Hence, the points $\{(-1/2-n,-1/2)\}_{n=0}^\infty$
and $\{(1/2+n,-1/2)\}_{n=0}^\infty$ will form the lattices for two
irreducible representations of $\{A^3,B^3\}$. A similar argument can
be established for the line defined by $c'=1/2$ and the operators $\{
A^4, B^4\}$. Therefore, we have constrained the original
$(a',c')$-plane from all the ${\bf R}^2$ into its more relevant part,
composed by two invariant interrelated sectors plus two invariant
lines.

\section{Schr\"odinger equations}

Let us start this section by introducing a free index operator $M$, written 
in the confluent hypergeometric notation as follows
\be
\label{map}
M\, f(a,c;x) := M_{(a,c)}f(a,c;x),\quad 
M_{(a,c)}:= Y_{(a,c)} \circ \varphi^{-1}(a,c;x)
\ee
where
\[
\varphi(a,c;x) =  \left[ \frac{e^x}{x^c \, \,y'(x)}
\right]^{1/2}
\]
and $Y_{(a,c)}$ is an auxiliary operator changing the independent variable 
$x$ into $y$
\[
Y_{(a,c)}: \left\{
\begin{array}{l}\, x \longrightarrow x(y), \\ [1ex]
 \, F(x) \longrightarrow \Phi(y) \equiv F(x(y))\ .
\end{array} 
\right.
\]
The action of the operator $M$ on the kernel elements of 
$L_{(a,c)}$ produces the mapping from the confluent hypergeometric equation 
$L_{(a,c)} f(a,c;x)=0$ to the time independent Schr\"{o}dinger equation
\be
\label{scro}
\left[ - \frac{d^2}{dy^2} + V(y) \right] \psi_s(y) = E \psi_s(y),
\ee
where $2m/\hbar^2=1$ and the label $s$ stands for the dependence on certain 
Schr\"odinger parameters. In other words, 
\be
f(a,c;x)\, \harr{M}{} \, \psi_s(y) =
\varphi^{-1} (a,c; x(y)) \, f(a,c; x(y)).
\ee
It is easy to check that the identification of $E$ and $V(y)$ in
(\ref{scro}) depends exclusively on the specific analytical
realization of $y'(x)$, {\it i.e.\/}, on the function we have taken
as the new independent variable $y(x)$. Once we know how to connect
the confluent hypergeometric equations to a class of Schr\"{o}dinger
equations, we can translate all the studied mathematical properties
into physical ones within the framework of the stationary
Schr\"{o}dinger wave-functions. Therefore, by using the first order
intertwiners of Table~1 and $M$, we can establish the following
commutative diagram
\be
\label{cuadro}
\begin{array}{ccc}
f(a,c;x)  \, & \, \harr{M}{}  \, &  \, \psi_s(y) \\
 & & \\ [-1ex]
\varr{{ A}^i_{(a,c)}}{} \varru{}{{ B}^i_{(\atil,\ctil)}} &
&\varru{{\bf B}^i_{\tilde s}}{}
\varr{}{{\bf A}^i_s} \\
 & & \\ [-1ex]
f(\atil, \ctil;x) & \harr{}{M} & \psi_{\tilde s}(y) 
\end{array}
\ee

\noindent
where the operators ${\bf A}^i_s$ and ${\bf B}^i_{\tilde s}$ 
relate Schr\"{o}dinger wave-functions with different labels:
\[
\begin{array}{l}
{\bf A}^i_s \, \psi_s(y) \equiv M { A}^i_{(a,c)} \, M^{-1} \,
\psi_s(y) 
\propto \psi_{\tilde s}(y),\\[2ex]
{\bf B}^i_{\tilde s} \, \psi_{\tilde s}(y) \equiv M {
B}^i_{(\atil,\ctil)} \, M^{-1} \, \psi_{\tilde s}(y) \propto \psi_{
s}(y). 
\end{array}
\]
Now, we proceed to identify the immediate physical potentials $V(y)$
related with the c.h.f. by means of the above mentioned
transformations. We shall focus on the main information drawing the
readers attention to our previous work \cite{Neg00} for details. In
each case $\ell$ stands for the angular quantum number $\ell \in {\bf
Z}^+$ and $E$ for the involved energy eigenvalues. Moreover, for all
the following examples, the c.h.f. $f(a',c';x)$ are understood as
appropriate linear combinations of the basis elements (\ref{basis}). 
Notice that it is always possible to choose between two different
combinations of the pair $(a',c')$, which are labeled by an $\pm$
subindex in each case. In addition, we have clearly established the
part of the $(a'_n, c'_m)$-lattice where the involved c.h.f.  live. 
The specific physically allowed values of these points can be
obtained by asking for the square integrability property of the
corresponding Schr\"odinger functions $\psi_s$. 

\noindent
{\bf N-dimensional harmonic oscillator}

\[
V^N_{\mbox \rm osc}(y) \equiv y^2 +
\frac{(2\ell +N -1)(2\ell + N -3)}{4 y^2}, \quad N \geq 2, \qquad
y=x^{1/2}
\]

\[
\left\{
\begin{array}{l}
\ell = {c_+'+1 -N/2} \geq 0 \\[1ex]
E= - 2 a_+' >0\\[1ex]
\mbox{\rm upper part of the L.I.S.}
\end{array}
\right.\qquad \qquad
\left\{
\begin{array}{l}
\ell = {1 - c_-' - N/2 } \geq 0 \\[1ex]
E= -2 a_-'>0\\[1ex]
\mbox{\rm lower part of the L.I.S.}
\end{array}
\right.
\qquad
\]

\noindent
{\bf One-dimensional harmonic oscillator}

\[
V_{osc}(y) =y^2, \qquad y=x^{1/2}
\]

\[
\left\{
\begin{array}{l}   
c_+' =1/2 \\ [1ex]
E=- 2 a_+'>0 \\[1ex]
\mbox{\rm upper left invariant line}
\end{array}
\right.\qquad \qquad
\left\{
\begin{array}{l}
c_-' =-1/2 \\ [1ex]
E= -2 a_-'>0 \\[1ex]
\mbox{\rm lower left invariant line}
\end{array}
\right.
\]

\noindent
{\bf N-dimensional Coulomb potential}

\[
V_{\rm Coul}^N(r) =-\frac{2}{y}+ \frac{(2\ell+N -1) (2\ell +N
-3)}{4y^2},
\qquad y \propto x
\]

\[
\left\{  
\begin{array}{l}
\ell = \frac{c_+'+2 -N}2\\ [1ex]
E= -\left(\frac{2}{a_+'} \right)^2 \\[1.5ex]
\mbox{\rm upper part of the L.I.S.}
\end{array}
\right.\qquad \qquad
\left\{
\begin{array}{l}
\ell = \frac{2 - c_-' - N }2\\ [1ex]
E= -\left(\frac{2}{a_-'} \right)^2 \\[1.5ex]
\mbox{\rm lower part of the L.I.S.}
\end{array}
\right.
\]

\noindent
{\bf Morse potential}

\[
V_M^{\lambda}(y) = \left( \frac{\alpha}{2} \right)^2
\left( e^{2\alpha y} -
2\lambda \, e^{\alpha y} \right), \quad y=(\ln x)^{1/\alpha}, 
\quad \alpha>0, \, \, \lambda>0
\]

\[
\left\{
\begin{array}{l}
c_+' = \frac2{\alpha} \sqrt{-E} \geq 1 \\ [1ex]
\lambda= -a_+' \\[1ex]
\mbox{\rm upper part of the L.I.S.}
\end{array}
\right. \qquad \qquad
\left\{  
\begin{array}{l}
c_-' = -\frac2{\alpha} \sqrt{-E} \leq 1\\ [1ex]
\lambda= - a_-' \\[1ex]
\mbox{\rm lower part of the L.I.S.}
\end{array}
\right.
\]

\noindent
As we can see, the same Schr\"odinger function $\psi_s$ can be
constructed by means of the c.h.f. parametrized by $(a'_+, c'_+)$ as
well as that with parameters $(a'_-, c'_-)$. As $c'_-$ is always such
that $c'_-=-c'_+$, these functions live in different parts of the
same (left) sector or line. Hence, it could be enough to consider
either the upper or the lower part of that sector or line to describe
the physical solutions of the related quantum problem. In other
words, there is a mapping 2:1 from the kernel ${\cal K}_{(a,c)}$ to
the Hilbert space ${\cal H}$ of square integrable functions $\psi_s$.
A situation leading to a representation of $Q$ in ${\cal H}$ as an
operator which does not transform the general solutions of the
Schr\"odinger equation but changing only its parametrization. The
same is true for the reflection operator $V$ developing the first
Kummer transformation on the c.h.f. 

In summary, the action of the operator $M$ on the c.h.f. (living only
in the afore mentioned sectors) gives the stationary Schr\"odinger
solutions for the oscillator, Coulomb, and Morse potentials. As we
have seen, $M$ connects the c.h.o. (\ref{uno}) with the Schr\"odinger
equation related with these potentials. Hence, only the left and
right invariant sectors, besides the invariant lines we are dealing
with, are the responsible for bound states in the physical problems
here considered (we will not discuss here the states corresponding to
the continuum spectrum of these examples.) This is why they have been
called the {\em physical sectors} of the confluent hypergeometric
functions space. On the other hand, the same mapping $M$ allows the
connection among the wave-functions of the diverese quantum systems
we have presented. As a very last example, let us consider the $N=2$
case for the oscillator and the Coulomb potentials. In order to fix
the notation, we first write the energy of the discrete spectrum as
follows:
\be
E_O = 2n_O,\quad E_C = \frac{-1}{(n_C +1/2)^2},\quad 
E_M = -\frac{\alpha^2}{4} \nu_M^2
\ee
where $O$, $C$ and $M$ mean oscillator, Coulomb, and Morse,
respectively. Thus, we have

\noindent
{\bf Oscillator-Morse}:
\be
\psi_M^{(\nu_M,\lambda_M)}(y) \propto 
e^{-\alpha y/4}\psi_O^{(n_O,\ell_O)}(e^{\alpha y/2}),\quad
n_O{=} \lambda_M , \ell_O {=}\nu_M.
\label{AB}
\ee

\noindent
{\bf Oscillator-Coulomb}:
\be
\psi_H^{(n_C,\ell_C)}(y) \propto 
y^{1/4}\psi_O^{(n_O,\ell_O)}(2y^{1/2}/(2n_O {+}1)^{1/2}),\quad
n_O {=} 2n_C {+}1, \ell_O{=} 2\ell_C.
\label{AC}
\ee
For other dimensions $N$ of the oscillator or Coulomb potentials, the
relations keep on valid, but with other restrictions on the values of
$(n,\ell)$.  The above formulas (\ref{AB})-(\ref{AC}) do not change
the number of nodes but they change the local density probability so
that there is a {\em mathematical mapping} among these three
potentials which does not represent a physical equivalence.

\vskip0.5cm

\noindent 
{\bf Acknowledgements:} 
JN and LMN aknowledge financial support by DGES projects (PB94--1115
and PB98--0370) from Ministerio de Educaci\'on y Cultura (Spain) and
also by Junta de Castilla y Le\'on (CO2/197 and CO2/199). ORO
acknowledges support by CONACyT project 32086E (Mexico).

\vskip0.5cm

\end{document}